\input{psfig}

\def\eV{\,{\rm eV}}

\def\sec{\,{\rm sec}}
\def\Gyr{\,{\rm Gyr}}

\def\Mpc{\,{\rm Mpc}}

\def\kms{\,{\rm km\,s^{-1}}}
\def\mpl{{m_{\rm Pl}}}

\def\la{\mathrel{\mathpalette\fun <}}

\def\fun#1#2{\lower3.6pt\vbox{\baselineskip0pt\lineskip.9pt
  \ialign{$\mathsurround=0pt#1\hfil##\hfil$\crcr#2\crcr\sim\crcr}}}

\documentstyle{article}
\begin{document}

\title{The Case for $\Lambda$CDM}

\author{Michael S. Turner\\
The University of Chicago and Fermilab}

\maketitle
\begin{abstract}

The case is simple:  there is no compelling theoretical
argument against a cosmological constant and
$\Lambda$CDM is the {\it only} CDM model that is consistent with
all present observations.  $\Lambda$CDM has two noteworthy
features:  it can be falsified in the near future (the prediction
$q_0\sim -{1\over 2}$ is
an especially good test), and, if correct, it has important
implications for fundamental physics.

\end{abstract}

\section{Introduction}

\subsection{Motivation}
Inflation is a bold and expansive idea that stands upon the tall
shoulders of the hot big-bang cosmology \cite{inflation1,inflation2}.
It holds the promise of
extending the standard cosmology to times as early as $10^{-32}\sec$,
of addressing almost all of the pressing issues in cosmology, and
of shedding light on the unification of the forces and particles of Nature.
Inflation is ripe for testing and the cold dark matter (CDM) theory
of structure formation plays a central role.

While there is no standard model of inflation, there
are three robust predictions:
(i) spatially flat Universe ($\Omega_0 \equiv \sum_i\Omega_i =1$
where $i=$ baryons, cold dark matter, hot dark matter, vacuum energy,
radiation, etc.); (ii) nearly scale-invariant spectrum of gaussian density
perturbations; and (iii) nearly
scale-invariant spectrum of gravitational waves.  (A few would
dispute the prediction of a flat Universe \cite{omeganotone1,omeganotone2};
however, I believe there is ample reason for calling it a robust
prediction \cite{inflation2}.)  These predictions,
together with the big-bang nucleosynthesis determination of the baryon density,
$\Omega_Bh^2 = 0.008 - 0.024$ \cite{cst}, and the failure of hot
dark matter models to account for the structure observed today \cite{nothdm},
make the cold dark matter theory of structure formation an
important secondary prediction.

When the CDM scenario emerged more than
a decade ago many referred to it as a ``no-parameter
theory'' because it was so specific compared to previous
models of structure formation \cite{faberetal}.  However,
this was enthusiasm speaking
as there are cosmological quantities that must be specified in any
theory of structure formation.  These parameters lead to families of CDM models.
Fortunately, observations are becoming good enough to both
decisively test inflation and discriminate between CDM models.

\subsection{CDM parameters}

Broadly speaking the parameters can be organized
into two groups \cite{dgt}.  First are the cosmological parameters:
the Hubble constant $h$; the density of
ordinary matter $\Omega_B h^2$; the power-law index $n$
that describes the shape of the spectrum of density perturbations and
the overall normalization of the power spectrum $A$ [$P(k)=Ak^n$];
the level of gravitational waves, specified by the
ratio of their contribution to the variance of the quadrupole anisotropy
to that of density perturbations ($\equiv T/S$) and the tensor
power-law index $n_T$.

The inflationary parameters
are in this group because there is no standard model for inflation.
They are related to the scalar-field potential
$V(\phi )$ that drives inflation
\begin{eqnarray}
S  \equiv {5\langle |a^S_{2m}|^2\rangle \over 4\pi} & = &
        {2.2 (V_*/\mpl^4) \over (\mpl V_*^\prime /V_*)^2} \\
T  \equiv {5\langle |a^T_{2m}|^2\rangle \over 4\pi} & = &
        0.61 (V_*/\mpl^4)\\
n-1 & = & -{\mpl \over 8\pi}\left( {V_*^\prime \over V_*}\right)^2
        +{\mpl^2\over 4\pi}\left( {V_*^\prime\over V_*}\right)^\prime\\
n_T  & = & -{\mpl \over 8\pi}\left( {V_*^\prime \over V_*}\right)^2
\end{eqnarray}
where prime indicates derivative with respect to $\phi$ and
$\phi_*$ is the value of the scalar field when scales of
order $10^4h^{-1}\,$Mpc crossed outside the horizon during inflation.
It should be noted that the scalar and tensor power spectra
are not exact power laws; e.g., the variation of the scalar index $n$
with scale (or running) is given by \cite{arthur}:
\begin{eqnarray}
{dn\over d\ln k}=
& - & {1\over 32\pi^2}\left({{m_{\rm Pl}}^3V^{\prime\prime\prime}\over V}\right)
\left({{m_{\rm Pl}} V^\prime\over V}\right)\nonumber \\
& + & {1\over 8\pi^2} \left({{m_{\rm Pl}}^2V^{\prime\prime}\over V}\right)
\left({{m_{\rm Pl}} V^\prime\over V}\right)^2
 -  {3\over 32\pi^2}\left({m_{\rm Pl}} {V^\prime\over V}\right)^4
\end{eqnarray}
I know of no model for which $n=1$ (exact scale invariance);
for several interesting models
$n\sim 0.94 - 0.96$; for natural or power-law inflation
$n$ can be as small as $0.7$; for hybrid inflation $n$ can
can be as large as 1.2, or as small as 0.7 \cite{models1,models2}.
The level of gravitational radiation can be negligible
or significant, and typically, $dn/d\ln k \simeq -10^{-3}$.

The second group of parameters specifies the composition of invisible matter
in the Universe:  radiation, dark matter, and cosmological
constant.  Radiation refers to relativistic
particles:  the photons in the CBR, three massless
neutrino species and possibly other undetected relativistic particles
(some particle-physics theories predict the existence of additional
massless particle species \cite{taucdm1,taucdm4}).  The level of radiation
today is important as it determines when the transition from
radiation domination to matter domination took place, and thereby
determines the characteristic turnover scale in the present CDM power spectrum,
$k_{\rm EQ}$ (Fig.~1).

Dark matter could include other particle relics in addition to CDM.
For example, a neutrino species of mass $5\eV$ (or two or
more neutrino species whose mass summed to $5\eV$)
would account for about 20\% of the critical density
($\Omega_\nu = m_\nu/90h^{2}\eV$).  Predictions for
neutrino masses range from $10^{-12}\eV$ to several MeV, and
there is some experimental evidence that at least one of
the neutrino species has mass \cite{numass1,numass2,numass3,numass4,numass5,numass6}.

In the modern context a cosmological constant corresponds
to an energy density associated with the quantum vacuum.  Since
there is no reliable calculation of the quantum vacuum energy \cite{cosmoconst},
the existence of a cosmological constant
must be regarded as a logical possibility.

\subsection{CDM family of models}

The original no-parameter CDM model, or standard CDM,
is characterized by simple choices for the
cosmological and the invisible matter parameters:
precisely scale-invariant density perturbations ($n=1$),
$h=0.5$, $\Omega_B =0.05$, $\Omega_{\rm CDM}=0.95$;
no radiation beyond the photons and the three massless neutrinos; no
dark matter beyond CDM; no gravitational radiation
and zero cosmological constant.  The overall normalization of
the power spectrum (i.e., $A$)
was determined by setting the {\it rms} mass fluctuation in spheres
of radius $8h^{-1}\,$Mpc ($\equiv \sigma_8$) equal to the inverse of
the bias parameter $b\sim 1 -2$, allowing for the likely possibility
that light (in the form of optically bright galaxies) is more
clustered than mass.

The COBE detection of CBR anisotropy on angular scales
of $10^\circ$ to $90^\circ$ changed the normalization procedure.
By requiring the predicted the level of CBR anisotropy to
be consistent with COBE, the power spectrum is
normalized on very-large scales
($10^3h^{-1}\Mpc$) without regard to biasing.  COBE also put the stake
through the heart of standard CDM:  COBE-normalized standard CDM
predicts too much power on
small scales \cite{jpo,ll} (see Fig.~1).  When normalizing CDM
by large-angle CBR anisotropy, the level of gravitational radiation
must be specified because some of the anisotropy on large
angular scales could arise from gravity waves:
a higher level of gravitational radiation leads
to a lower level of density perturbations.

The standard CDM set of parameters is not sacred; it was simply a
starting point.  In making the case for $\Lambda$CDM I
will discuss four ``families'' of CDM models.  They are distinguished
by their invisible matter content: standard invisible matter content (sCDM);
extra radiation ($\tau$CDM); small hot dark matter component ($\nu$CDM);
and a cosmological constant ($\Lambda$CDM).  There are, of course, more
complicated possibilities, e.g., $\nu\Lambda$CDM, etc.

\section{Evidence:  observations favor $\Lambda$CDM}

In reviewing the observations I will show that of these
four models only $\Lambda$CDM is consistent with all present data.

\subsection{CBR anisotropy}
There are now more than ten
independent detections of CBR anisotropy on angular scales
from $0.5^\circ$ to $90^\circ$ and the angular power spectrum
is beginning to show a Doppler peak at $l\sim 200$ (as expected
for a flat Universe).  However, the strongest constraints come from the
COBE four-year data set \cite{dmr4yra,dmr4yrb,dmr4yrc} which
implies:  quadrupole
anisotropy $Q = (18 \pm 2 \pm 1)\, \mu$K (for $n=1$), where the
error is statistical + systematic arising from galaxy subtraction, and
$n=1.1\pm 0.2$.  CBR anisotropy serves to normalize the
power spectrum and exclude models with $n< 0.7$.

\subsection{Power spectrum}
There are three robust
constraints to the power spectrum from observations of
the contemporary Universe:  the shape
derived from several redshift surveys \cite{pd} (Fig.~1);
the value of $\sigma_8$ derived from the abundance of x-ray clusters
\cite{sigma8cluster}, $\sigma_8 = (0.5-0.8)\Omega_{\rm Matter}^{-0.56}$;
the level of inhomogeneity on small scales ($\sim 0.2h^{-1}\Mpc$)
required to insure early structure
formation (neutral gas in damped Ly-$\alpha$ systems at redshift
four, $\Omega_{\rm DLy-\alpha}h = 0.001 \pm 0.0002$ \cite{Dlya1,Dlya2}).
For all four families, there are a variety of cosmological parameters
for which the COBE-normalized power spectrum is consistent with
these three constraints (Fig.~2).

\subsection{Matter density}
Determinations of $\Omega_0$ and $\Omega_{\rm Matter}$
provide powerful tests of inflation as well as discriminating
between the different CDM models.   At present, the best the strongest
statements that can be made are:  $\Omega_0 \ge \Omega_{\rm Matter}
> 0.3$, based upon the peculiar velocities of thousands of
galaxies (including the Milky Way) \cite{pecvel1,pecvel2}; $\Omega_\Lambda
< 0.7$, based upon the frequency of gravitational lensing of QSOs
\cite{qsolensing}.  While a decisive determination of $\Omega_0$
is lacking, there is little
evidence to suggest that $\Omega_{\rm Matter} = 1$ -- as predicted
in all but $\Lambda$CDM -- and much evidence that $\Omega_{\rm Matter}
\sim 0.3$ -- as predicted by $\Lambda$CDM.

\subsection{Hubble constant/Age of Universe}
Together, these two fundamental cosmological parameters
have great leverage.  Determinations of
the Hubble constant based upon a variety of techniques (Type Ia and II
supernovae, IR Tully-Fisher and fundamental plane methods)
have converged on a value between $60\kms\Mpc^{-1}$ and $80\kms\Mpc^{-1}$.
This corresponds to an expansion age of less than $11\Gyr$ for a flat,
matter-dominated model; for $\Lambda$CDM, the expansion age can be
significantly higher, as large as $16\Gyr$ for $\Omega_\Lambda = 0.6$ (Fig.~3).
On the other hand, the
ages of the oldest globular clusters indicate that the Universe is
between $13\Gyr$ and $17\Gyr$ old; further, these age determinations,
together with the
those for the oldest white dwarfs and the long-lived radioactive elements,
provide an
ironclad case for a Universe that is at least $10\Gyr$ old
\cite{age1,age2,age3}.
{\it Unless the age of the Universe and the Hubble constant are near the
lowest values consistent with current measurements, only $\Lambda$CDM
model is viable.}

\subsection{Cluster baryon fraction}  Clusters are large enough
that the baryon fraction should reflect its universal value,
$\Omega_B/\Omega_{\rm Matter} = (0.008 - 0.024)h^{-2}/(1-\Omega_\Lambda )$.
Most of the (observed) baryons in clusters are in the hot,
intracluster x-ray emitting gas.  From x-ray measurements of
the flux and temperature of the gas, baryon fractions
in the range $(0.04 - 0.10)h^{-3/2}$ have been
inferred \cite{gasratio1,gasratio2,gasratio3}; further,
a recent detailed analysis and comparison to numerical models
of clusters in CDM indicates
an even smaller scatter, $(0.07\pm 0.007)h^{-3/2}$ \cite{evrard}.
From the cluster baryon fraction and $\Omega_B$,
$\Omega_{\rm Matter}$ can be inferred:  $\Omega_{\rm Matter}
= (0.25\pm 0.15)h^{-1/2}$, which for the lowest Hubble constant
consistent with current determinations ($h=0.6$) implies
$\Omega_{\rm Matter} = 0.32 \pm 0.2$.  {\it Unless one of the assumptions
underlying this analysis is wrong, only $\Lambda$CDM is viable.}

\subsection{And the winner is ...}

Since only $\Lambda$CDM is consistent with all the observations
there can be little debate that it is the current strawman
for structure formation.  (Unless one is willing to
dispute some of the observations or their interpretations --
e.g., the Hubble constant, age of the Universe, or cluster baryon
fraction.)  Further, taken together, the constraints argue
for $\Omega_\Lambda
\sim 0.5 - 0.65$ and $h\sim 0.6 - 0.7$ as the best fit model (Fig.~4).  Others
have reached similar conclusions \cite{bestfit1,bestfit2}.

\section{Falsification of $\Lambda$CDM}

At the moment, the case for $\Lambda$CDM hinges upon the cluster
baryon fraction and measurements of the age and Hubble constant.
However, in the near future there are a number of
tests that can distinguish $\Lambda$CDM from its CDM siblings.

\begin{itemize}

\item Deceleration parameter.  This is the most striking test:  $\Lambda$CDM
predicts $q_0 \equiv {1\over 2} - {3\over 2}\Omega_\Lambda\sim
-{1\over 2}$, while the other CDM models predict $q_0 = {1\over 2}$.
Two groups (The Supernova Cosmology Project and The High-z Supernova Team)
are hoping to determine $q_0$ to a precision of $\pm0.2$ by
using distant Type Ia supernovae ($z\sim 0.3 - 0.7$) as standard candles.
Together, they discovered more than 40 high redshift supernovae
last fall and winter and both groups should be announcing results soon.

\item Hubble constant.  Since the Universe is at least $10\Gyr$ old,
a determination that the Hubble constant is $65\kms\Mpc^{-1}$ or greater
would rule out all models but $\Lambda$CDM; on the other hand,
a determination that
the Hubble constant is below $55\kms\Mpc^{-1}$ would undermine much of
the motivation for $\Lambda$CDM.

\item Cluster baryon fraction.  This strongly favors
$\Lambda$CDM.  Further evidence that x-ray measurements
have correctly determined the total cluster mass (e.g., from weak
gravitational lensing) and baryon mass (e.g., from AXAF) would
strengthen the case for $\Lambda$CDM.  On the other hand,
discovery of a systematic effect
that lowers the cluster baryon fraction by a factor of two
(e.g., underestimation of cluster mass because gas is not supported
by thermal pressure alone, or overestimation of cluster gas mass
because the gas is clumped) would undermine the case for $\Lambda$CDM.

\item Gravitational lensing.  It has long been appreciated that
$\Lambda$CDM predicts a much higher frequency of gravitationally lensed
QSOs \cite{edturner1,edturner2}; however, modelling uncertainties have
precluded setting a limit more stringent than $\Omega_\Lambda \la 0.7$
\cite{qsolensing}.
With new QSO lensing surveys coming
gravitational lensing should not be forgotten as a striking
signature of $\Lambda$CDM.

\item Early structure formation.  Because $\Lambda$CDM is slightly
antibiased (Fig.~1) structure formation commences earlier.
The study of the Universe at high redshift by HST and Keck
will test this prediction.

\item Redshift surveys.  The differences
in the level of biasing, power spectrum and redshift space distortions
between $\Lambda$CDM and the other models are significant.  The
two large redshift surveys coming on line (SDSS and 2dF) should
be able to discriminate between the different CDM models.

\item Theory.  The theoretical underpinnings of $\Lambda$CDM could be changed
by new arguments against or in favor of a cosmological constant.

\item CBR anisotropy in the MAP/COBRAS/SAMBA era.  The high-resolution
CBR maps that will be made by these two satellite-borne experiments
will settle the issue decisively (Fig.~5) -- among other things, by
determining both $\Omega_0$ and $\Omega_\Lambda$ to better than
10\% \cite{learncbr}.

\end{itemize}

\section{Final Remarks}

Inflation is a bold, expansive and attractive extension of
the standard cosmology, and the cold dark matter theory of structure
formation provides a crucial test of it.  Although I am not
wedded to any CDM model -- I will be happy to see
any one proven correct -- the only model consistent with all
present observations is $\Lambda$CDM.  To be fair, the case hinges upon
the Hubble constant and cluster baryon fraction, neither one of which
has been settled completely; however, new observations (especially
$q_0$) should clarify matters soon.

To end, I summarize my view of the world models discussed by this
panel.  $\Lambda$CDM is the best fit to
the present data; sCDM is the most elegant;
$\nu$CDM has the most striking signature -- around
$5\eV$ of neutrino mass; the defect models are the most
interesting; and OCDM is my worst nightmare!

\section*{Bibliographic Notes} 

\begin{figure}[t]
\centerline{\psfig{figure=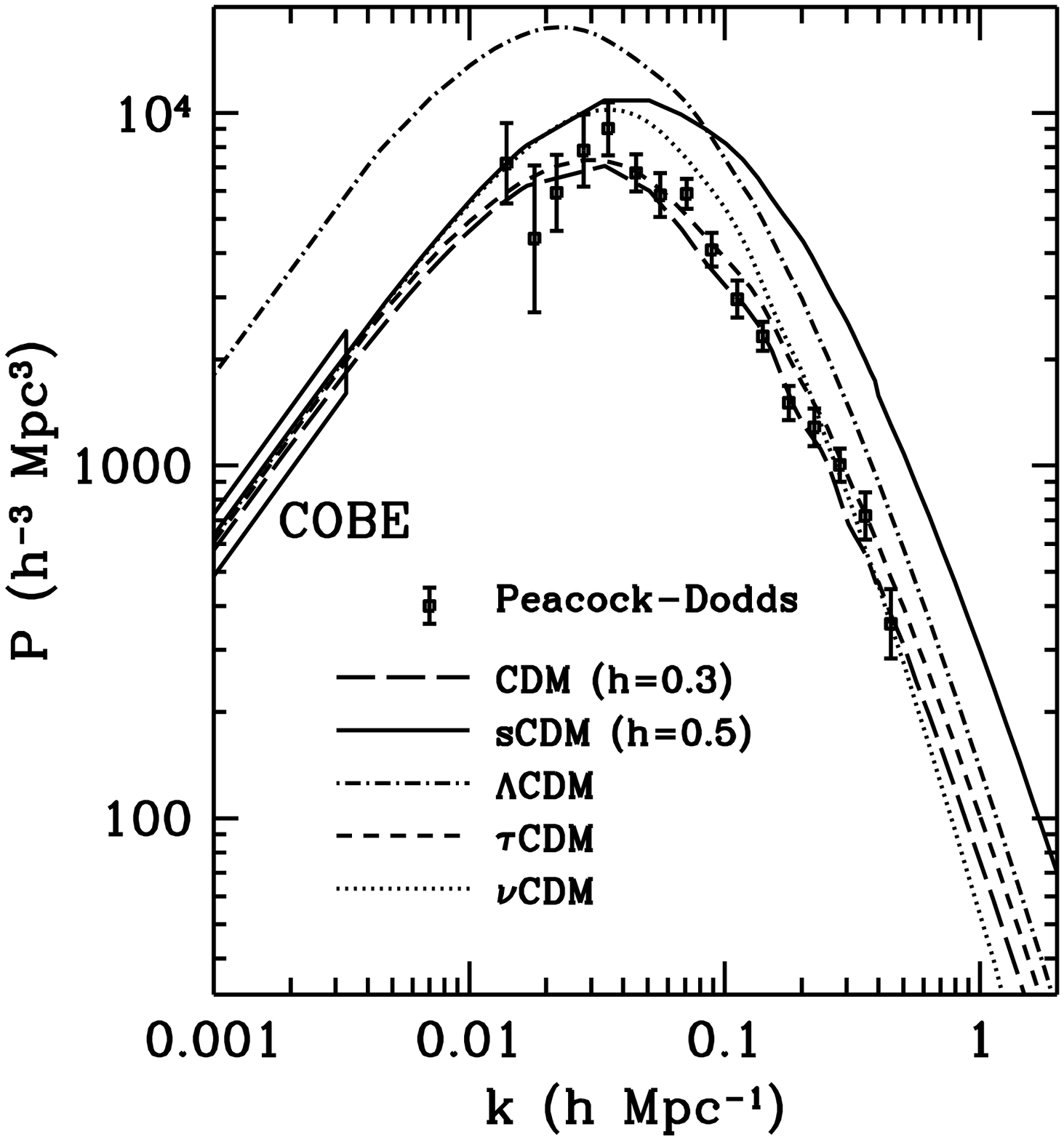,width=3.5in}}
\caption{Measurements of the power spectrum,
$P(k) = \langle |\delta_k|^2\rangle = Ak^n$, and the predictions of different
COBE-normalized CDM models.
The points are from several redshift surveys compiled by Peacock and Dodds
\protect\cite{pd}; the models are:  $\Lambda$CDM
with $\Omega_\Lambda =0.6$ and $h=0.65$; standard CDM (sCDM),
CDM with $h=0.35$; $\tau$CDM (with the energy equivalent
of 12 massless neutrino species) and $\nu$CDM with $\Omega_\nu = 0.2$
(unspecified parameters have their standard CDM values).
The offset between a model and the points indicates the level of biasing.
Note, $\Lambda$CDM does not pass through the COBE rectangle because
a cosmological constant alters the relation between the power spectrum
and CBR anisotropy (from Ref.~\protect\cite{dgt}).}
\end{figure}

\begin{figure}[t]
\centerline{\psfig{figure=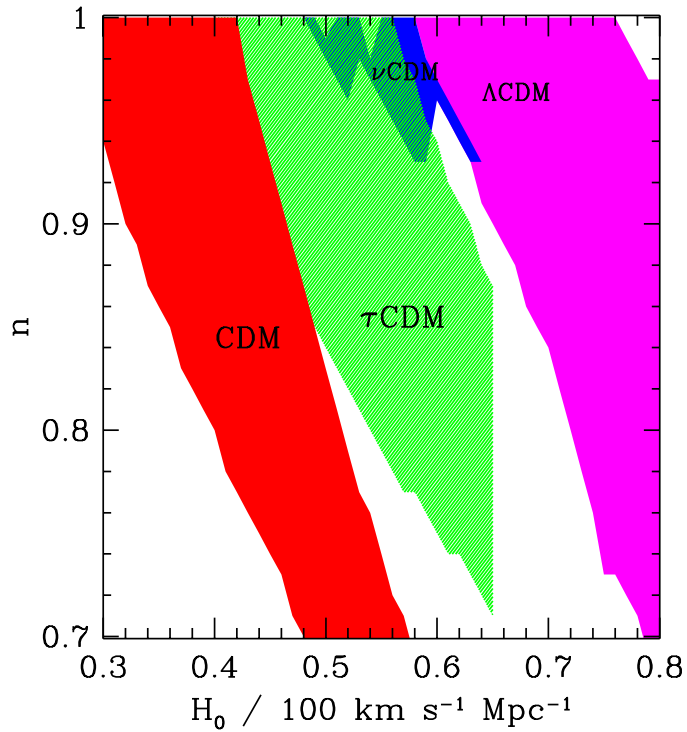,width=3.5in}}
\caption{Acceptable values of the cosmological
parameters $n$ and $h$ for CDM models with standard
invisible-matter content (CDM), with 20\% hot dark matter ($\nu$CDM),
with additional relativistic particles (the energy equivalent
of 12 massless neutrino species, denoted $\tau$CDM), and with a cosmological
constant that accounts for 60\% of the critical density ($\Lambda$CDM).
The $\tau$CDM models have been truncated at a Hubble constant
of $65\kms\Mpc^{-1}$ because a larger value would result in a
Universe that is younger than $10\Gyr$ (from Ref.~\protect\cite{dgt}).}
\end{figure}

\begin{figure}[t]
\centerline{\psfig{figure=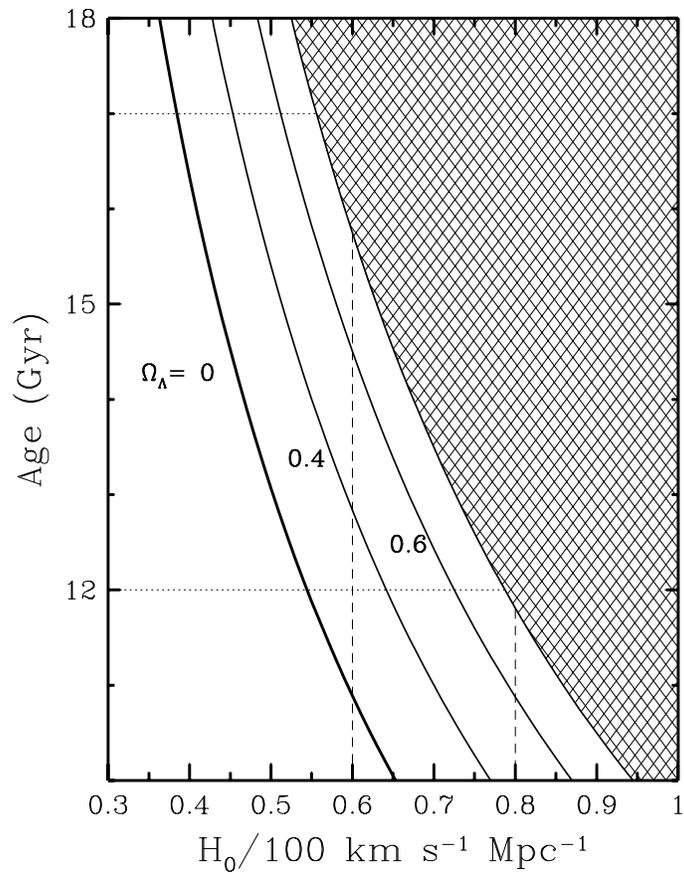,width=3.5in}}
\caption{The relationship between age and $H_0$ for flat-universe
models with $\Omega_{\rm Matter} = 1 - \Omega_\Lambda$.
The cross-hatched region is ruled out because
$\Omega_{\rm Matter} < 0.3$.  The dotted lines indicate
the favored range for $H_0$ and for the age of the Universe
(from Ref.~\protect\cite{dgt}).}
\end{figure}

\begin{figure}[t]
\centerline{\psfig{figure=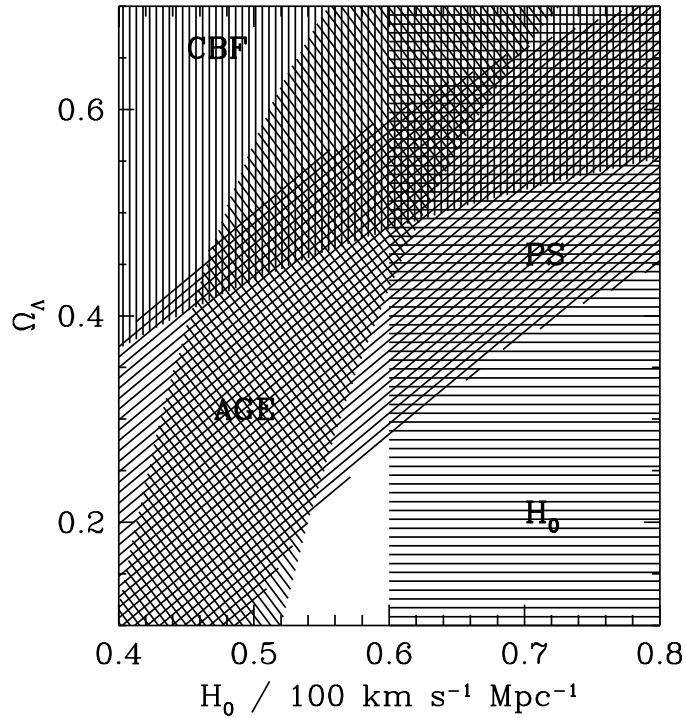,width=3.5in}}
\caption{Summary of constraints projected onto
the $H_0$ -- $\Omega_{\rm Matter}$ plane:
(CBF) comes from combining the BBN limit to the
baryon density with x-ray observations
of clusters; (PS) arises from the power spectrum;
(AGE) is based on age determinations of the Universe;
($H_0$) indicates the range currently favored for the
Hubble constant.  (Note the constraint $\Omega_\Lambda <0.7$
has been implicitly taken into account since the
$\Omega_\Lambda$ axis extends only to 0.7.)
The darkest region indicates the parameters
allowed by all constraints (from \protect\cite{bestfit1}.)}
\end{figure}

\begin{figure}[t]
\centerline{\psfig{figure=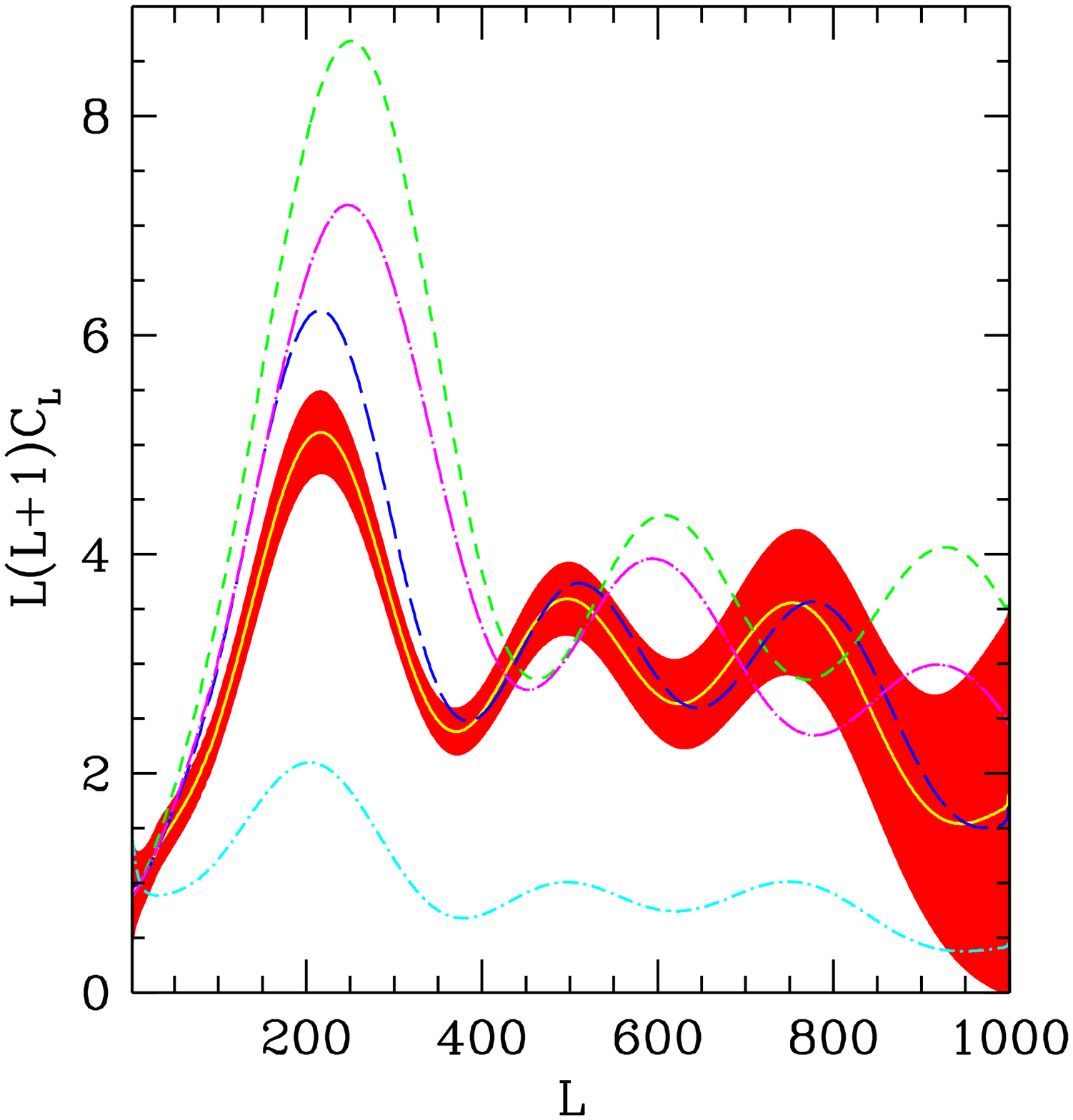,width=3.5in}}
\caption{Angular power spectra of
CBR anisotropy for several CDM models and the anticipated
uncertainty (per multipole) from a CBR satellite experiment
similar to MAP.   From top to bottom the CDM models
are:  CDM with $h=0.35$, $\tau$CDM with the energy equivalent
of 12 massless neutrino species, $\Lambda$CDM with $h=0.65$ and
$\Omega_\Lambda = 0.6$, $\nu$CDM with $\Omega_\nu = 0.2$,
and CDM with $n=0.7$ (from Ref.~\protect\cite{dgt}).}
\end{figure}


\begin{thebibliography}{inflation1}


\bibitem{numass2} Athanassopoulos, C. et al 1995, PhysRevLett, 75, 2650



\bibitem{numass6} Becker-Szendy, R. et al 1992, PhysRevD, 46, 3720

\bibitem{dmr4yra} Bennett, C.L. et al.1996, ApJLett, 464, L1

\bibitem{faberetal} Blumenthal, G. et al 1984, Nature, 311, 517

\bibitem{age1}  Bolte, M. \& Hogan, C.J. 1995, Nature, 376, 399

\bibitem{gasratio2} Briel, U.G. et al. 1992, A\&A, 259, L31

\bibitem{omeganotone1} Bucher, M., Goldhaber, A.S. \& Turok, N. 1995,
PhysRevD, 52, 3314

\bibitem{age2} Chaboyer, B., Demarque, P., Kernan, P.J. \& and Krauss, L.M.
1996, Science, 271, 957

\bibitem{cst}  Copi, C., Schramm, D.N. \& Turner, M.S. 1995, Science,
267, 192

\bibitem{age3} Cowan, J., Thieleman, F. \& Truran, J. 1991, ARAA, 29, 447

\bibitem{pecvel2} Dekel, A. 1994, ARAA, 32, 319

\bibitem{dgt} Dodelson, S., Gates, E. \& Turner, M.S. 1996, Science, in press.

\bibitem{taucdm1} Dodelson, S., Gyuk, G. \& Turner, M.S. 1994, PhysRevLett, 72, 3578

\bibitem{evrard} Evrard, A. 1996, in preparation

\bibitem{numass5} Fukuda, Y., et al 1994, PhysLettB, 335, 237

\bibitem{edturner2} Fukugita, M. \& Turner, E.L. 1991, MNRAS, 253, 99

\bibitem{dmr4yrb}  Gorski, K.M., et al.1996, ApJLett,464, L11

\bibitem{omeganotone2} Gott, J.R. 1992, Nature, 295, 304

\bibitem{inflation1} Guth, A.H. 1981, PhysRevD, 23, 347

\bibitem{numass3} Hill, J.E. 1995, PhysRevLett, 75, 2654

\bibitem{numass4} Hirata, K.S. et al. 1992, PhysLettB, 280, 146

\bibitem{learncbr} Jungman, G.,  Kamionkowski, M., Kosowsky, A. \& Spergel, D.
1996, PhysRevD, in press

\bibitem{taucdm4} Kikuchi, H. \& Ma, E. 1995, PhysRevD, 51, R296

\bibitem{qsolensing}  Kochanek, C.S. 1996, ApJ, in press

\bibitem{inflation2}  Kolb, E.W. \& Turner, M.S. 1990, The Early Universe
(Addison-Wesley, Redwood City, CA), chapt. 8

\bibitem{arthur} Kosowsky, A. \& Turner, M.S. 1995, PhysRevD, 52, R1739

\bibitem{bestfit1} Krauss, L. \& Turner, M.S. 1995, GenRelGrav, 27, 1137

\bibitem{Dlya1} Lanzetta, K., Wolfe, A.M. \& Turnshek, D.A. 1995, ApJ, 440, 435

\bibitem{ll} Liddle, A. \& Lyth, D. 1993, PhysRepts, 231, 1

\bibitem{models2} Lyth, D. \& Stewart, E. 1996, hep-th/9606412

\bibitem{jpo} Ostriker, J.P. 1993, ARAA, 31, 689

\bibitem{bestfit2} Ostriker, J.P. \& Steinhardt, P.J. 1995, Nature, 377, 600

\bibitem{numass1}  Parke, S. 1995, PhysRevLett, 74, 839

\bibitem{pd}  Peacock, J. \& Dodds, S. 1994, MNRAS, 267, 1020

\bibitem{Dlya2} Storrie-Lombardi, L.J.,  McMahon, R.G., Irwin, M.J., \& Hazard,
C. 1996, MNRAS, in press (astro-ph/9608147)

\bibitem{pecvel1} Strauss, M. \& Willick, J. 1995, PhysRepts, 261, 271

\bibitem{edturner1} Turner, E.L. ApJLett, 365, L43

\bibitem{models1} Turner, M.S. 1993, PhysRevD, 48, 3502

\bibitem{cosmoconst}  Weinberg, S. 1989, RevModPhys, 61, 1

\bibitem{gasratio3} White, D.A. \& Fabian, A.C. 1995, MNRAS, 273, 72

\bibitem{dmr4yrc} White, M. \& Bunn, E.F. 1995, ApJ, 450, 477

\bibitem{sigma8cluster}  White, S.D.M., Efstathiou, G. \& Frenk, C.S. 1993,
MNRAS, 262, 1023

\bibitem{gasratio1} White, S.D.M. et al. 1993, Nature, 366, 429

\bibitem{nothdm} White, S.D.M., Frenk, C. \& Davis, M. 1993, ApJLett, 274, L1

\end{thebibliography}
\end{document}